\documentclass[conference]{IEEEtran}
\IEEEoverridecommandlockouts
\usepackage{cite}
\usepackage{amsmath,amssymb,amsfonts}
\usepackage{algorithmic}
\usepackage{graphicx}
\usepackage{textcomp}
\usepackage{xcolor}
\def\BibTeX{{\rm B\kern-.05em{\sc i\kern-.025em b}\kern-.08em
    T\kern-.1667em\lower.7ex\hbox{E}\kern-.125emX}}
\begin{document}

\title{Supporting Evolution of Safety Cases within \\the Software Development Process} 

\title{Leveraging Traceability to Integrate Safety Analysis Artifacts into the Software Development Process}

\author{
\IEEEauthorblockN{Ankit Agrawal}
\IEEEauthorblockA{\textit{Department of Computer Science } \\
\textit{Saint Louis University}\\
Saint Louis, MO, USA \\
ankit.agrawal.1@slu.edu}
\and
\IEEEauthorblockN{Jane Cleland-Huang}
\IEEEauthorblockA{\textit{Department of Computer Science} \\
\textit{University of Notre Dame}\\
SouthBend, IN, USA \\
janehuang@nd.edu}
}

\maketitle

\begin{abstract}
Safety-critical system's failure or malfunction can cause loss of human lives or damage to the physical environment; therefore, continuous safety assessment is crucial for such systems. In many domains this includes the use of Safety assurance cases (SACs) as a structured argument that the system is safe for use. SACs can be challenging to maintain during system evolution due to the disconnect between the safety analysis and system development process. Further, safety analysts often lack domain knowledge and tool support to evaluate the SAC. We propose a solution that leverages software traceability to connect relevant system artifacts to safety analysis models, and then uses these connections to visualize the change. We elicit design rationales for system changes to help safety stakeholders analyze the impact of system changes on safety. We present new traceability techniques for closer integration of the safety analysis and system development process, and illustrate the viability of our approach using examples from a cyber-physical system that deploys Unmanned Aerial Vehicles for emergency response.

\end{abstract}


\section{Introduction}
Safety-critical systems are systems whose failure could result in loss of life, significant damage to the environment, or significant financial loss \cite{DBLP:conf/ifip13/GreenwellSK04}. Such systems must be developed systematically and rigorously. Given a set of requirements describing the system's functionality, we need to assure that associated hazards have been identified and appropriately addressed, typically using techniques such as  Fault-Tree Analysis (FTA) and Failure-Mode Effect and Criticality Analysis (FMECA). Beyond these techniques, it is increasingly common for organizations to construct claim-based safety arguments \cite{Hawkins2013} in the form of a  Safety Assurance Case (SAC).  A SAC decomposes high-level safety goals or claims into layers of arguments supported by safety evidence such as test-cases logs, simulation results, or formal proofs \cite{Bishop1998}, often using either the Claims-Arguments-Evidence notation \cite{netkachova2015tool} or the Goal Structuring Notation \cite{Kelly2004}. 

SACs are recommended, or even required, in many safety critical domains (e.g.,\cite{USFDA2014}); however, in a study by Cheng et al., safety experts reported that there are `no effective mechanisms for managing change' for a SAC, and that `SAC creation and maintenance has not been fully integrated into the software development process' \cite{DBLP:journals/corr/abs-1803-08097}. System change analysis becomes difficult because of insufficient domain-knowledge among safety stakeholders and the sheer complexity involved in safety-critical products. During change analysis, the safety stakeholders seek answers to questions such as  (1) why the system has changed, (2) what risk does this change mitigates, and (3) how this change can impact safety\cite{gleirscher2017arguing}. Further, safety stakeholders typically design and maintain SACs while the development team produces artifacts, such as mitigating requirements and test results, on which the SAC's safety arguments depend. These distinct roles also create a gap between system development and SAC maintenance processes. Therefore, it is crucial not only to establish traceable links between various safety artifacts such as Fault Tree, SAC,  and development artifacts such as Design Decisions and code, but to keep safety analysts informed of how changes in development artifacts may affect safety by documenting the rationales for changes to these development artifacts.

In this paper, we propose a solution to establish traceability links between various safety artifacts such as SACs and Fault Trees, and supplementing the SACs with rationales for changes in the development artifacts. Our goal is to improve maintainability of SACs and keep safety analysts informed of the rationale for changes in the development artifacts as the software evolves.  Our solution first utilizes Safety Artifact Forest Analysis (SAFA) \cite{DBLP:conf/icse/AgrawalKVRCL19,DBLP:conf/kbse/RodriguezNDC22,DBLP:journals/software/Cleland-HuangAV21}, which detects changes in the software development artifacts between the two versions of the underlying system and automatically generates vizualizations enabling analyst to easily navigate through changes. We establish traceable links between these auto-generated vizualizations and safety artifacts such as FTAs and SACs. Secondly, we discuss strategies to capture the reasons behind changes in software development artifacts and maintain change rationale information as part of the development artifacts. Finally, we demonstrate how this comprehensive traceability across multiple safety artifacts, supported by rationales for changes in the development artifacts, enables us to analyze the impact of changes on safety as the system evolves. To illustrate our approach, we provide examples from the DroneResponse system \cite{cleland2020requirements, agrawal2020next, cleland2020human}, which utilizes Unmanned Aerial Vehicles (UAVs) to support emergency response. Finally, we outline the open challenges ahead in a preliminary roadmap.

\section{System Artifacts Traceability}
\label{sec:artifacts}

\begin{figure*}[ht]
    \centering
    \includegraphics[width=0.97\textwidth]{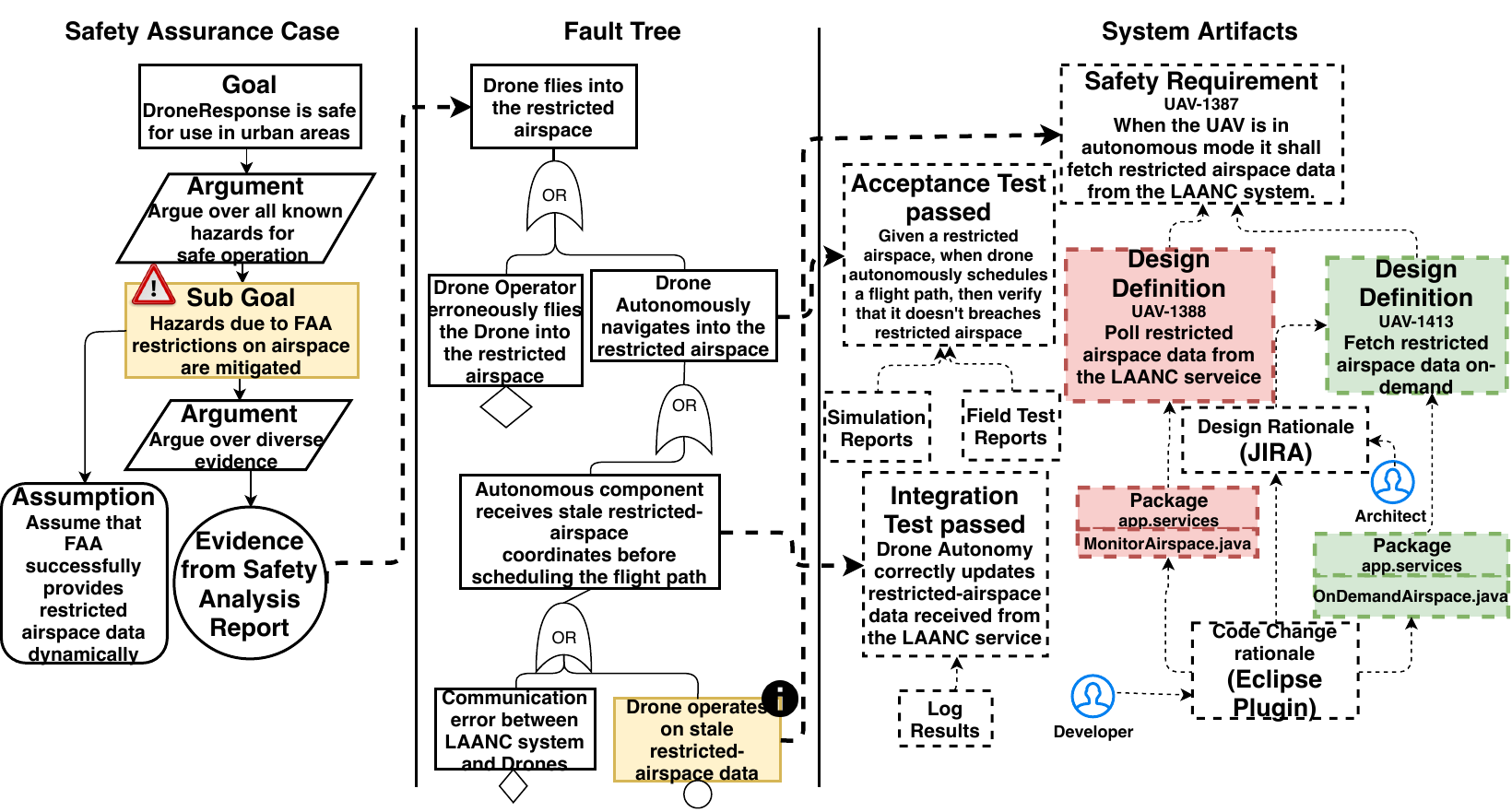}
    \caption{Software artifacts, fault tree nodes, and evidence nodes from the SAC are connected via trace links to integrate system development and safety processes and to provide support for managing the impact of change upon system safety.}
  
    \label{fig:SAC}
\end{figure*}

The SAFA framework retrieves artifacts from project repositories such as DOORS, Jira, and Github. Given a root node, such as a system level requirement, it constructs a vertical slice through the system according to the trace links defined in a Traceability Information Model (TIM). SAFA refers to such a tree as an Artifact Tree (AT). SAFA can compare a current version of the AT against an earlier baseline version to produce a Delta Tree (DT) which visualizes changes in the system. 
The right lane of Figure \ref{fig:SAC} shows a partial DT. Additional examples are provided in our prior work \cite{DBLP:conf/icse/AgrawalKVRCL19}. 

SAFA  detects additions (green), deletions (red), and modifications (blue) for requirements, design, code, tests, operating context, environmental assumptions, and other system artifacts. The delta tree visualization helps project stakeholders to identify changes and to investigate their impact on system safety.  It highlights these changes and recommends areas in which inspection is needed. Safety experts expressed the importance of providing rationales for each change \cite{DBLP:conf/icse/AgrawalKVRCL19}. While SAFA currently uses basic static analysis techniques to explain refactoring changes; richer rationales are needed that describe changes in requirements and design, as well as modifications to the code.  

We illustrate the proposed solution with examples from publicly available requirements dataset for DroneResponsecyber-physical system that deploys cohorts of UAVs to support emergency response mission such as search-and-rescue  \cite{DBLP:conf/icse/Cleland-HuangVB18}. The use of on-board intelligence by UAVs for autonomous navigation through airspace is one of the key requirements of multi-UAV autonomous systems. However, the US Federal Aviation Authority (FAA) regulates airspace usage and defines special-use or restricted-airspace where UAVs are not allowed to fly. Autonomous UAVs entering the restricted-airspace could cause accidents with commercial flights, military operations, or medivac deliveries.  Therefore, we consider flights into prohibited space to be a severe operational risk and have carefully designed mitigations into our system. 

The safety requirement UAV-1387 in Figure \ref{fig:SAC} states that ``When the UAV is in autonomous mode it shall fetch restricted airspace data from the LAANC system.'' In the previous version of the system, this was partially addressed by requiring the UAV to continuously check for airspace information while in autonomous mode (Design Definition - UAV 1388). However, in the new system, this requirement was replaced by conducting a more economical check when new flight paths are planned (Design Definition - UAV 1413). The delta tree, produced by SAFA, clearly highlights this design change, showing the replacement of Design Definition UAV-1388 and its associated code in red, and the inclusion of Design Definition UAV-1413 and its associated code in green.

\section{Integrating System Artifacts with Safety Assets}
\label{sec_safetyassets}
When developing a safety-critical system, a preliminary hazard analysis (PHA) is performed \cite{Leveson1995} to identify high-level hazards that represent undesirable states of the system.  Each of these hazards is then explored through an associated Fault Tree (FT) \cite{Storey:1996:SCC:524721} or a FMECA model \cite{DBLP:journals/ansoft/LutzW97}. In this paper, we illustrate our approach using FTs; however, our techniques are also applicable to FMECAs. Fault Tree Analysis (FTA) is a top-down approach that starts by analyzing the high-level risk and then uses boolean logic to depict a chain of events causing system-level risk. The middle lane of Figure \ref{fig:SAC} shows a partial FT for risks associated with UAV flights in restricted airspace. 


FTs and FMECAs are typically used as part of a SAC's argumentation structure to show that a specific fault has been sufficiently mitigated.  Therefore a link should be established from a FT to its relevant argument in the SAC.
An example of a SAC argument using GSN is depicted in the left lane of Figure \ref{fig:SAC}. This SAC uses a complete fault tree as evidence to support the claim that  \textit{hazards due to FAA restrictions on airspace are mitigated}. To maintain horizontal traceability (depicted by horizontal dashed arrows in Figure \ref{fig:SAC}), an explicit link is established between an evidence node in the SAC and the root node of the FT. In turn, links are established from multiple nodes in the FT to sub-trees of system artifacts. For example, the intermediate fault node of \emph{`drone autonomously navigates into the restricted airspace'} is linked to an acceptance test as verification that the fault has been mitigated. Similarly, one of the contributing basic faults \emph{`Drone operates on stale restricted-airspace data'} is linked to the previously discussed safety requirement (UAV-1387).

In our solution, we establish traceability links between system artifacts, safety assets (e.g., FTs and FMECAs), and SACs to propagate changes back-and-forth between safety assets and system artifacts. Figure \ref{fig:SAC} illustrates that a single change in the design of the system could trigger and propagate notifications and warnings across the linked safety artifacts (yellow nodes).

\section{Capturing Rationales}
\label{sec:Rationales}
To support Safety Analysts in analyzing the impact of changes in system artifacts on the overall safety of the system, our approach strategically captures rationales from developers and other project stakeholders as they make changes that impact artifacts linked directly or indirectly to an FT, FMECA, or other safety assets. For example, in the FT depicted in the middle lane of Figure \ref{fig:SAC}, we observe that the leaf node describing the basic fault \emph{`Drone operates on stale restricted airspace data'} is mitigated through safety requirement UAV-1387, which states that \emph{`When the UAV is in autonomous mode it shall fetch restricted airspace data from the LAANC system.'}  As indicated by the red nodes, the original design satisfied this requirement through continuously fetching restricted airspace data during flight, while the current version (shown in green) replaces this functionality with a single fetch each time a new flight path is planned. A safety analyst will need to determine whether this change adversely impacts safety.  

\begin{figure}[htbp]
    \centering
    \includegraphics[width=0.90\columnwidth]{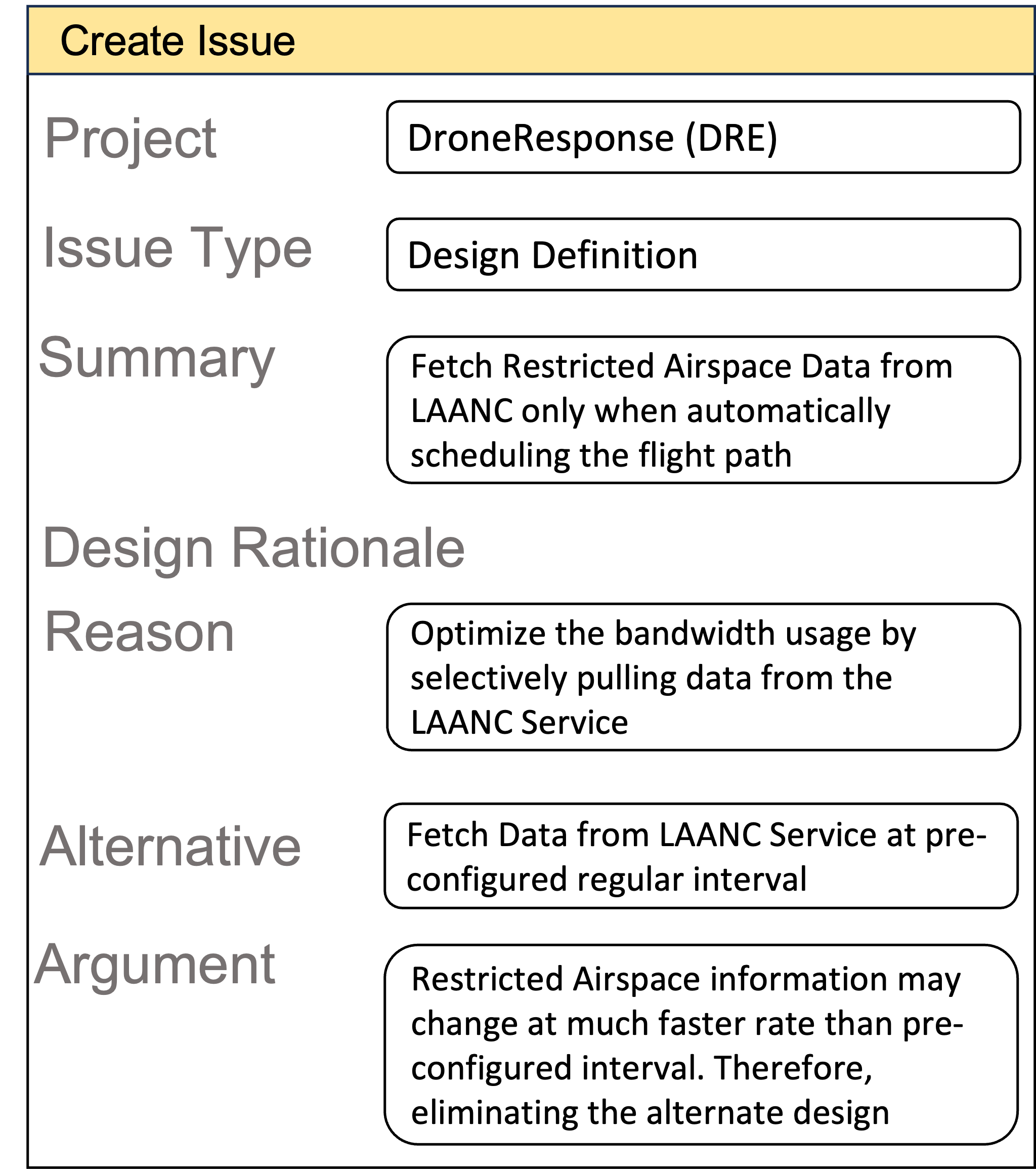}
    \caption{User interface for capturing design decision details, including reasons, alternatives, and arguments.}
    \label{fig:design-rationale}
\end{figure}

Burge et al suggested capturing reasons, alternative options, and arguments to provide rationales for design decisions \cite{Burge1}. We therefore elicit a rationale for any design decision that links directly to an FTA. In the case of DroneResponse, all requirements and design decisions are captured in Jira, and Figure \ref{fig:design-rationale} shows possible instrumentation of the Jira environment to elicit design rationales when the original design requirement (UAV-1388) is replaced by a new one (UAV-1413).  


\begin{figure}[htbp]
    \centering
    \includegraphics[width=0.90\columnwidth]{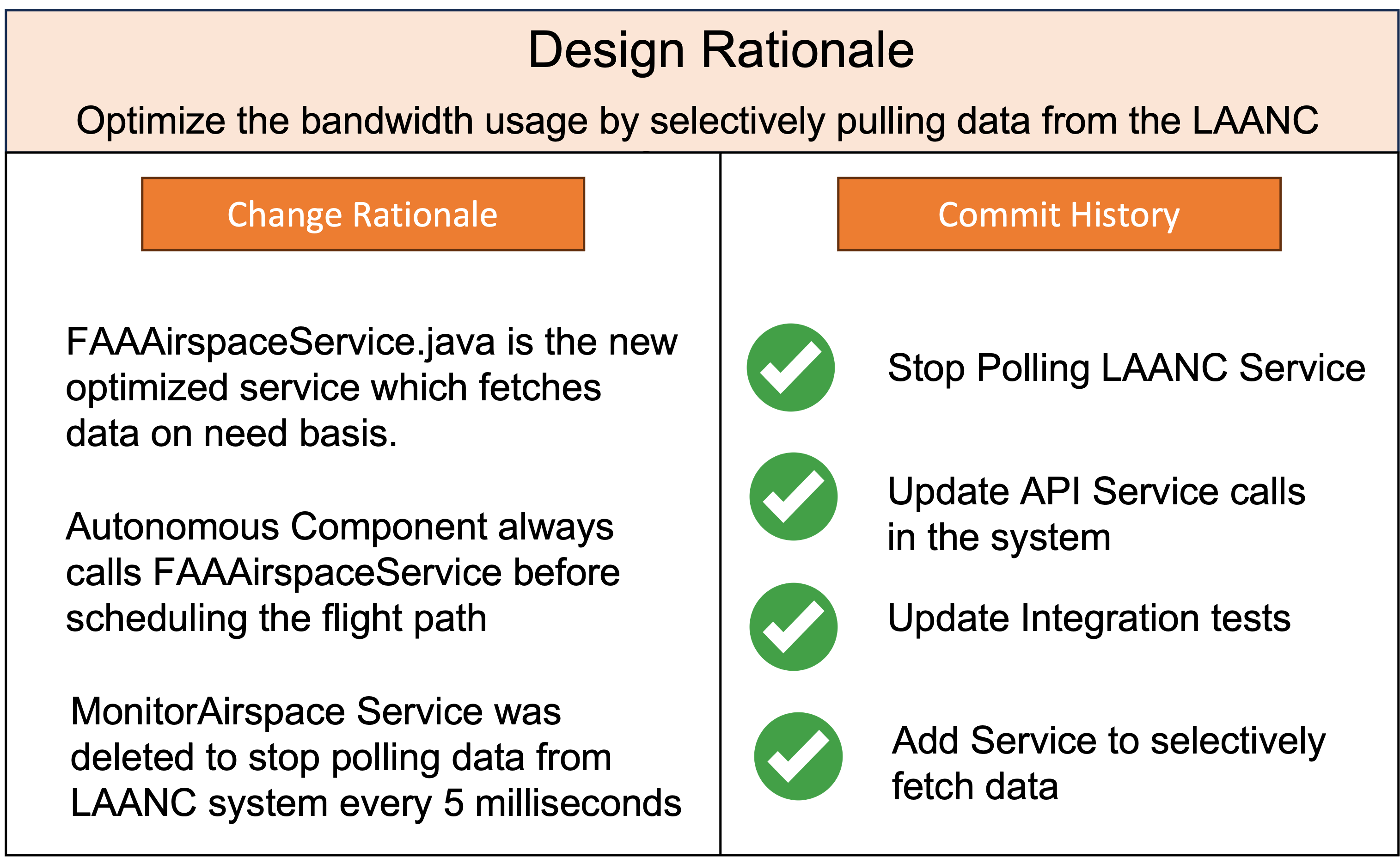}
    \caption{Eclipse Plugin User Interface for capturing code-level change rationale}
    \label{fig:code-rationale}
\end{figure}

At the code level we capture change rationales. These are akin to commit messages, but at the granularity of each modified class instead of an entire change set. In our example, the MonitorAirspace.java file is replaced by a new optimized service OnDemandAirspace.java that fetches data on-demand. Figure \ref{fig:code-rationale} shows the user interface of a prototype IDE plugin that could be used to capture granular details of the change. The plugin not only elicits a justification and explanation of the change from the developer, but also visualizes the contribution that the class makes to an FTA (e.g., Fig. \ref{fig:SAC}) and provides links to prior commit messages and previous change justifications. 

The rationales captured as part of the change process provide crucial domain knowledge to conduct a thorough analysis of system changes on safety. Therefore, the domain knowledge acquired from these rationales, alongside visual representations of change impact, as depiceted in Figure \ref{fig:SAC}, can aid safety analysts to determine (1) whether current changes impact safety or not, (2) whether additional mitigations are needed, and/or (3) whether the FT, FMECA, and/or the SAC need to be updated to reflect new  hazards or new safety arguments.  Decisions made by safety analysts can then be propagated back to the development team in order to close the loop.



\section{Future Challenges}
\label{sec:RoadMap}
\begin{itemize}
    \item {\bf Knowledge Management:} As previously reported, the innate complexity of safety-critical systems means that experts from diverse disciplines are needed to construct and maintain a SAC \cite{DBLP:journals/corr/abs-1803-08097}. While we integrate basic rationale capture into SAFA, open questions include (1) What types of domain knowledge are needed by safety analysts to evaluate and/or construct a SAC? (2) How and when should this information be collected? (3) How can it be effectively used to support Safety Analysts? 
    \item {\bf Intelligent analysis of change: } Changes are introduced into a system at many different levels to accommodate changes in the environment, introduce functional enhancements, improve system qualities such as performance, reliability, or maintainability, or to correct errors. From a safety perspective we need to differentiate between harmful and non-harmful errors. Our current approach visualizes all change, captures stakeholders' change rationales, and provides basic explanations for change.  However, future systems should be able to leverage AI solutions to (1) analyze individual and composite changes in order to identify and explain patterns of change, (2) differentiate between harmless changes and those with potential safety impact, and (3) recommend remedial actions when safety is impacted.
    \item{\bf Tool Supported Integrated Environments} Our proposed approach requires trace links to be established across heterogeneous artifacts stored in diverse tools and repositories. Safety experts have previously reported that the lack of tool support and clear guidance make SAC creation and maintenance challenging \cite{DBLP:journals/corr/abs-1803-08097}. Open challenges therefore include (1) defining best practices for creating end-to-end traceability across safety assets and development artifacts, (2) providing integrated tool-supported environments for retrieving diverse artifacts and establishing effective  traceability that supports safety analysis, and (3) developing interactive visualization tools that display SACs, FTAs, FMECAs, rationales, and artifacts in ways that provide appropriate support for different types of users such as safety analysts and developers.
\end{itemize}

This paper has presented a preview of our current research  in addressing the disconnect between safety analysis and the software development process. As proposed in this paper, we are developing plugins to capture change rationales at the design and code level, and extending SAFA to show rationales in Delta view and link FTAs, FMECAs, and SACs in it.


\bibliographystyle{abbrv}
\bibliography{main}
\end{document}